\documentclass[12pt]{article}
\usepackage{amssymb,amsmath,eucal}
\begin{document}

\begin{titlepage}

                            \begin{center}
                            \vspace*{5mm}
\Large\bf{Almost \ additive \ entropy}\\

                            \vspace{2.5cm}

              \normalsize\sf    NIKOS \  \  KALOGEROPOULOS $^\dagger$\\

                            \vspace{2mm}
                            
 \normalsize\sf Weill Cornell Medical College in Qatar\\
 Education City,  P.O.  Box 24144\\
 Doha, Qatar\\

                            \end{center}

                            \vspace{2.5cm}

                     \centerline{\normalsize\bf Abstract}
                     
                           \vspace{3mm}
                     
\normalsize\rm\setlength{\baselineskip}{18pt} 

\noindent We explore consequences of a hyperbolic metric induced by the composition property 
of the Harvda-Charvat/Dar\'{o}czy/Cressie-Read/Tsallis entropy. 
We address the special case of systems described by small deviations of the non-extensive parameter
\ $q\approx 1$ \ from the ``ordinary" additive case which is described by the 
Boltzmann/Gibbs/Shannon entropy. By applying the Gromov/Ruh theorem for almost flat manifolds, 
we show that such systems have a power-law rate of expansion of their configuration/phase space volume.  
We explore the possible physical significance of some geometric and topological results of this approach.
                           
                             \vfill

                             
                             \vfill

\noindent\rule{8cm}{0.2mm}\\
   \noindent \small\rm $^\dagger$  E-mail: \ \  \small\rm nik2011@qatar-med.cornell.edu\\

\end{titlepage}


                                                                                 \newpage

 \normalsize\rm\setlength{\baselineskip}{18pt}

                     \centerline{\large\sc 1. \  \  Introduction}

                                                                                              \vspace{5mm}

The Harvda-Charvat [1], Dar\'{o}czy [2], Cressie-Read [3], [4] Tsallis [5], [6] entropy, henceforth to be called just Tsallis entropy for 
brevity and following the current nomenclature of non-extensive Statistical Mechanics, is a functional defined on the space of 
probability distributions \ $\rho$ \ of a system, given by
\begin{equation}
            S_q [\rho] \ = \ k_B \ \frac{1}{q-1} \ \left\{ 1 - \int_{\Omega} \ [\rho(x)]^q \ d\mu_{\Omega} \right\}
\end{equation}
where \ $k_B$ \ stands of the Boltzmann constant and \ $q\in\mathbb{R}$ \ is called the 
entropic or non-extensive parameter and measures, in a way, the degree of non-extensivity of the system. Here \ $\Omega$ \ 
stands for the space where the evolution of the system is described. Examples of such $\Omega$ are the configuration or 
phase space of a Hamiltonian system. There \ $d\mu_{\Omega}$ \ an appropriate measure defined on \ $\Omega$. \ 
Such a measure could be  the Liouville measure of a Hamiltonian system or the Sinai-Ruelle-Bowen measure of a dissipative 
system, for instance.  It is straightforward to verify that 
\begin{equation}
          \lim_{q\rightarrow 1} \ = \ S_{BGS} 
\end{equation}
where \ $S_{BGS} $ \ indicates the Boltzmann/Gibbs/Shannon (BGS) entropy 
\begin{equation}
         S_{BGS} [\rho]  \  =  \ - k_B \int_{\Omega} \ \rho(x)  \log \rho(x) \ d\mu_{\Omega} 
\end{equation}  
Henceforth we will be setting \ $k_B = 1$ \ for simplicity.\\

When we compare the axioms that can be used to define the BGS [7] - [9] and the Tsallis [10], [11] entropies, we see that  the only 
substantial difference between these entropic forms  is the way they compose for stochastically independent 
(sub)systems. Let \ $\Omega_1$ \ and \ $\Omega_2$  \ be two such subsystems and let \ $\Omega_1 + \Omega_2$ \ indicate the 
system that arises from their interaction. Then stochastic independence amounts to multiplication of the marginal probability distributions 
describing them 
\begin{equation}
     \rho_{\Omega_1 + \Omega_2} \ = \ \rho_{\Omega_1} \cdot \rho_{\Omega_2}
\end{equation} 
For such stochastically independent subsets, the BGS entropy is additive 
\begin{equation} 
        S_{BGS} [\rho_{\Omega_1 + \Omega_2}] \ = S_{BGS} [\rho_{\Omega_1}] + S_{BGS} [\rho_{\Omega_2}]
\end{equation}
but, by contrast, the Tsallis entropy obeys 
\begin{equation}
        S_q [\rho_{\Omega_1 + \Omega_2}] \  = \ S_q [\rho_{\Omega_1}] + S_q [\rho_{\Omega_2}] + (1-q) S_q [\rho_{\Omega_1}] 
                                                                                           S_q [\rho_{\Omega_2}]
\end{equation}
This Tsallis entropy composition property (6) naturally leads to the definition of a generalized addition [12], [13]
\begin{equation}
          x \oplus_q y = x + y + (1-q) xy
\end{equation}
so that (6) can be recast in a more ``natural" form as 
\begin{equation}
          S_q [\rho_{\Omega_1 + \Omega_2}] \  = \ S_q [\rho_{\Omega_1}] \oplus_q S_q [\rho_{\Omega_2}] 
\end{equation}
In effect, the difference between the BGS and the Tsallis entropies, at this formal level, amounts to understanding the difference 
between the ordinary and the generalised addition (7). In order to be able to work with convenient algebraic structures, a generalised 
product, distributive with respect to the generalised addition was defined, independently of each other, in [14] and [15]. This gave rise 
to a deformation of the reals, indicated by \ $\mathbb{R}_q$ \ [15]. \ Therefore a comparison between the ordinary addition and (7) 
also amoubnted  to a comparison between \ $\mathbb{R}$ \ and \ $\mathbb{R}_q$. \\

It may be worth noticing that \ $\mathbb{R}$ \ and \ $\mathbb{R}_q$ \ are algebraically indistinguishable.  We were able to 
construct an explicit  field isomorphism between them  \ $\tau_q: \mathbb{R} \rightarrow \mathbb{R}_q$ \  for $q\in [0,1)$  \ 
which was given by [15]
\begin{equation}
         \tau_q (x) \ = \ \frac{(2-q)^x - 1}{1- q}
\end{equation}
Given this algebraic identification, one way to compare \ $\mathbb{R}$ \ and \ $\mathbb{R}_q$ \ in a meaninglful way  
that may give rise to non-trival results  is through their metric properties [16]. 
Since all metrics on \ $\mathbb{R}$ \ are isometric, up to re-parametrization, and the instrinsic geometry of $\mathbb{R}$ is rather trivial, 
the simplest set on which a comparison between \ $\mathbb{R}$ \ and \ $\mathbb{R}_q$ \ could give any sought after non-trivial results is 
\ $\mathbb{R}^2$ \ (topologically), or a bit more accurately, from a metric viewpoint, \ $\mathbb{R} \times \mathbb{R}_q$. \ Such a 
metric comparison within the Riemannian category, which is essentially a comparison between the ordinary addition and (7), 
gave rise to the hyperbolic metric [15], [16], expressed in Cartesian \ $(x,y)$ \ coordinates as
\begin{equation}  
        {\bf g}_h \ = \ \left( 
                                 \begin{array}{ll}
                                        1 & 0 \\
                                        0 & e^{-2tx}
                                 \end{array}
                              \right)             
\end{equation}
where 
\begin{equation}
        t \   = \  \log (2-q)
\end{equation} 
It turned out that (10) has constant negative sectional curvature 
\begin{equation}
      k \ = \ - \left\{ \log (2-q) \right\}^2 
\end{equation}
so it is a hyperbolic metric, up to a global re-scaling. 
For comparison purposes, in this metric formalism, the ordinary addition gives rise to the ordinary (``Euclidean")  metric
\begin{equation}
      {\bf g}_E \ = \ \left( 
                                 \begin{array}{ll}
                                        1 & 0 \\
                                        0 & 1
                                 \end{array}
                              \right)             
\end{equation}
whose sectional curvature, obviously, vanishes, i.e. it is flat. So, in a way, the Tsallis entropy is a ``hyperbolic counterpart"
of the BGS ``Euclidean" entropy. After establishing (10) through a solvable group construction, we have more recently examined 
some of its implications [17] - [19].\\

An alternative way to compare the ordinary addition and (7) was proposed in [20], [21]. This way utilises in an essential manner the
multiplication properties of matrices and reduces the comparison of between the ordinary and the generalised additions 
to the algebraic and geometric properties of an ambient linear matrix group. Although most properties of this proposed embedding 
have not been  sustantianlly expored yet, but at least it is known that the ambient space turns out to be the 3-dimensional Heisenberg 
group. Several implications of this construction for the Tsallis entropy composition are examined in [21], one of which is the power-law 
growth rate of volumes in the configuration/phase space. 
Such a behaviour had been previously discussed for discrete (binary) variables in [22], [23].
This power-law may be an important prediction of the Tsallis entropy in that it may provide a feature that distinguishes between 
systems that are described by the BGS and the Tsallis entropies. 
It is our understanding that even such a statement is somewhat tenuous at the time of this 
writing, and for this reason we would not even entertain the idea of speculating on the validity of a possible converse statement.\\

The main goal of the present work is to establish power-law growth rate of the volumes of phase space for systems described by the Tsallis 
entropy using a different approach than the one of [20], [21]. We make several comments, on the side, pertaining to potential implications of 
this formalism for systems that are described by the Tsallis entropy. The necessity of our alternative approach to that of [20], [21]
 is dictated by the requirement of robustness of the constructions of [20], [21]. 
In these works there was an embedding of the ordinary and generalised additions into 
what turned out to be the Heisenberg group. As it occurs for any such embedding, one can ask on whether, or to what extent,  the final 
results depend on it. This question is quite typical in geometry but acquires an especially important role in Physics, where physical results 
cannot possibly depend on the formalism that is used to attain them. 
Hence an alternative way to derive results helps increase our confidence  
in that the outcomes may have physical significance as opposed to being potentially irrelevant illusions of formal manipulations.
We will use the Riemannian formalism alluded to above [15]-[19], as opposed to the Carnot-Carath\'{e}odory approach  of [20], [21].
Our approach relies in a most crucial manner to the Gromov-Ruh theorem on almost flat manifolds, so the current work can be seen as 
an application of that  result to the case of Tsallis entropy and, by extension, to non-extensive Statistical Mechanics. 
It should be noted that a limiting feature of the current approach is that it can only address systems that are close to being additive, in the 
usual sense of the word, corresponding to \ $q\approx 1$. \ Hence we will be only covering cases where the ordinary additivity is almost 
obeyed. Section 2 contains the main body of the paper where some background material and basic references are also provided. 
Section 3 briefly recaps with some conclusions and provides  comments and speculation that may be addessed in future work. \\

                                                                                             \vspace{10mm}
  
 
                                                     \centerline{\large\sc 2. \  \  Almost \ flatness  }
                                                     
                                                                                                   \vspace{5mm}
                                                                                                           
\noindent {\bf 2.1} \ It is may be worth observing that the difference between the ordinary addition and (7) becomes very pronounced for 
large  values of the entropies, namely when \ $|x| \gg 1$ \ and \ $|y| \gg 1$. \ Naturally, what is considered a small or a large entropy largely 
depends on the system under consideration.  Given this, one can see that a distinction between the BGS and Tsallis entropies for given 
system would be more evident, if one compares the most highly entropic objects of the system. In this sense, the Tsallis entropy is a 
``non-linear deformation" of the BGS entropy where the effect of the deformation becomes much more pronounced at the 
``boundary at infinity" of the configuration/phase space of the model at hand. This corresponds to values of \  $S_q \rightarrow +\infty$. \
For such values (7) is much more akin to ordinary multiplication rather than ordinary addition. This is quantified precisely by the 
field isomorphism  (9) which is essentially an exponential map. \\ 


\noindent {\bf 2.2} \ Second, the construction leading to (10) can be readily generalised to \ $\mathbb{R}^n$, \ by considering the warped 
product on \  $\mathbb{R} \times \mathbb{R}^{n-1}$ \ given in the Cartesian coordinates \ $(x_0, x_1, \ldots x_{n-1})$ \ by  
\begin{equation}
      {\bf g}_h \ = \ \mathrm{diag} (1, e^{-2tx_0}, \ \ldots, \ e^{-2tx_0})       
\end{equation}
where all the $n-1$ diagonal elements are equal, except the first one, and where  all off-diagonal components of the metric 
tensor \ ${\bf g}_h$ \ vanish.\\

Consider now the case in which \ $q \approx 1$. \ This is the case at the ``interface" of validity of the BGS ($q=1$) and Tsallis \ ($q\neq 1$) \
entropies where the usual additivity is violated and (7) is obeyed, but not by much. Given that \ $q\in [0,1)$, \ we can write 
\begin{equation}
        q = 1 - \varepsilon, \hspace{10mm} 0 < \varepsilon \ll 1
\end{equation}
which allows us to rewrite (7) as
\begin{equation}
        x\oplus_q y \ = \ x + y + \varepsilon xy
\end{equation}
In terms of \ $\varepsilon$, \ (11) simplifies, in the lowest non-trivial order approximation, to 
\begin{equation}
     t \ = \ \varepsilon
\end{equation}
and 
\begin{equation}
      k \ = \   -\varepsilon^2
\end{equation}
This states that the induced metric from the Tsallis entropy composition property (7) in the almost additive limit (15) is almost flat, 
a term that we subsequently explain.\\ 


\noindent {\bf 2.3} \ As is well-known [24], [25], the Riemann tensor is a nonlinear expression involving the metric, its inverse and their  
derivatives up to second order. To be more concrete, the Riemann tensor \ $R: TM \otimes TM \otimes TM \rightarrow TM$ \ 
on an $n$-dimensional manifold \ $M$ \ whose tangent bundle is indicated by
\ $TM$ \ and with vector fields \ $X, Y, Z \in TM$ \ is given in terms of the Levi-Civita connection \ $\nabla$ \ by 
\begin{equation}
     R(X,Y) Z \ \equiv \ \nabla_X \nabla_Y Z - \nabla_Y\nabla_X Z - \nabla_{[X,Y]} Z
\end{equation}
Its components are  expressed in the coordinate basis \ $(\partial_1, \ldots \ \partial_n ) \in TM$ \ and its dual \ 
$(dx^1, \ \ldots, \ dx^n) \in T^\ast M$, \ in one convention, as
\begin{equation}
   R_{ijkl} \ = \  {\bf g} (\partial_i, R(\partial_k, \partial_l) \partial_j)
\end{equation}
where \ ${\bf g}$ \ stands for the metric tensor on \ $M$. \ Indices are raised and lowered by using ${\bf g}$. This expresses in 
the given coordinate systems the bijection between \ $TM$ \ and \ $T^\ast M$ \ which is provided by the metric tensor \ ${\bf g}$. \  
In terms of the metric, rather than the Levi-Civita connection \ $\nabla$, \ the above expression can be recast in terms of the 
Christoffel symbols \ $\Gamma$ \ as 
\begin{equation}
    \nabla_{\partial_i} \ \partial_j \ = \ \Gamma^k _{\hspace*{1mm} ij} \ \partial_k 
\end{equation}
giving
\begin{equation}
      \Gamma^i _{\hspace*{1mm} jk} \ = \ \frac{1}{2} {\bf g}^{im} (\partial_j {\bf g}_{mk} + \partial_k {\bf g}_{jm} - \partial_m {\bf g}_{jk} )
\end{equation}
which gives for the components of the Riemann tensor
\begin{equation}
    R^i _{\hspace*{1mm} jkl} \ = \ \partial_k \Gamma^i _{\hspace*{1mm} lj} - \partial_l \Gamma^i _{\hspace*{1mm} kj} 
                                                                     + \Gamma^m  _{\hspace*{3mm} lj} \  \Gamma^i _{\hspace*{1mm} km} 
                                                                             - \Gamma^m _{\hspace*{3mm} kj} \ \Gamma^i  _{\hspace{1mm} lm}
\end{equation}
The summation convention over repeated indices ranging from \ $1$ \ to \ $n = dim M$ \ is assumed in (21)-(23).
The sectional curvature \ $k_x (\partial_i, \partial_j)$ \ is defined as the Gaussian curvature of the 2-plane spanned by 
\ $\partial_i$ \ and \ $\partial_j$ \ at $x\in M$, up to normalisation  
\begin{equation}
        k_x(\partial_i, \partial_j) \ = \ \frac{ {\bf g} (\partial_i, R(\partial_i, \partial_j)\partial_j)}{{\bf g}(\partial_i, \partial_i) 
                                                                       {\bf g}(\partial_j, \partial_j) - [{\bf g} (\partial_i, \partial_j)]^2}
\end{equation}
As such, $k$ is an element of the Grassmann bundle \ $G_{2,n}(M)$.\\

From these definitions it is obvious that when the distance function scales by a factor of \ $t$, \  the Riemann tensor and the 
sectional curvature scale by \ $t^{-2}$. \  This can be easily seen in the case of a sphere of radius \ $r$ \ whose curvature 
scales as \ $r^{-2}$. \  Curvature quantifies the intuitive concept that a large radius sphere has smaller curvature as it turns 
its tangent vectors ``less abruptly", whereas a smaller curvature sphere turns them ``more abruptly", hence it has a larger curvature.
The morale of all this, is that if we want to re-scale a metric on \ $M$ \ and still capture some of its geometric features, 
one way is to  demand that the product \ $k$ \ times (linear dimension)$^2$  to be constant or at least to obey some bounds. 
One of the simplest such relations that takes into account the inevitable sectional curvature variations is to demand that     
\begin{equation} 
         ( \sup_{x\in M} |k|) \ D^2 \ < \ \delta
\end{equation}
where \ $0 < \delta \ll 1$ \ and where \ $D$ \ is the diameter of \ $M$ 
\begin{equation}
      D = \sup_{x,y \in M}  d(x,y)
\end{equation}
Manifolds obeying (26) are called almost flat [26]. We see from (18) the Tsallis entropy in the almost additive regime \ 
$q \approx 1$ \ is to induce the almost flat metric  (10) in the place of the flat metric (13) induced by the BGS entropy.\\


\noindent {\bf 2.4} \ Consider a group \ $G$ \ which may be discrete or continuous. It acts upon itself by left multiplication.
Let \ $g, h \in G$ \ and consider the adjoint action
\begin{equation}
      Ad_g (h) \ \equiv \  [g, h] \ \equiv \ g^{-1} h^{-1} g h
\end{equation}
Let $A$, $B$ be subgroups of $G$ and indicate by $[A, B]$ the group commutator of $A$ and $B$, namely the subgroup of \ 
$G$ \ which is generated by the commutators $[a,b]$ with $a\in A$ and $b\in B$
\begin{equation}
      [A, B]  \ = \ \langle a^{-1}b^{-1}ab, \ \ \ \forall \ a\in A, \ \forall \ b\in B \rangle
\end{equation}
Define the consecutive commutators 
\begin{equation}
G_1 = G, \ \ G_2 = [G, G], \ \ \ldots, \ \ G_{n+1} = [G_n, G], \ \ \ \ \ \ \  \ n\in\mathbb{N} 
\end{equation}
These subgroups can be arranged in the following lower central series 
\begin{equation}
  G_1 > G_2 > \ldots > G_n > G_{n+1} > \ldots 
\end{equation}
A group is called nilpotent if the lower central series terminates after a finite number of steps. When $G_{n+1}$ is trivial then 
$G$ is called $n$-step nilpotent. As examples: an Abelian group 
is nilpotent since $G_2$ is trivial. Hence an Abelian group is $1$-step nilpotent. The Heisenberg group has $G_3$ trivial 
hence it is $2$-step nilpotent. More generally, for the case of Lie groups of finite dimension, all of which are realised as matrix 
groups, according to Ado's theorem: an \ $n\times n$ \ matrix group is nilpotent if it is upper triangular and if all the elements of 
its diagonal are equal to the unit. Then it turns out the this matrix group is ($n-1$)-step nilpotent. A nil-manifold is a homogeneous 
space diffeomorphic to the quotient of \ $G/H$ \ where \ $G$ \ is a nilpotent Lie group and \ $H$ \ is a  subgroup of \ $G$. \\


\noindent {\bf 2.5} \ The central result used in the present work is a theorem proved by Gromov in [26] and further strengthened 
by Ruh in [28]. The proof of [26] provides a vast generalisation of the comparatively older and better known theorem of Bieberbach. 
The latter theorem states, in the compact category any flat manifold is a quotient of a torus by a discrete group. 
In a similar spirit, but with a far more general set of techniques involved, Gromov proves [26] that an almost flat manifold has a finite cover 
which is diffeomorphic to a nil-manifold.  Moreover Gromov proved that every nil-manifold admits an almost flat metric. 
The specifics of the statements and the proofs of [26] were explained in considerable detail in [27] to which we refer for further 
explanations and details.  Subsequently, in [28], Ruh strengthened the main result of [26] by proving that an 
almost flat manifold itself and not just a finite cover of it, possess a nilpotent structure. Hence an almost flat manifold has a nilpotent 
structure so it is a nilmanifold. Moreover, any nilmanifold is the quotient of a simply connected nilpotent group by a discrete cocompact 
subgroup.\\

 As noted in subsection {\bf 2.4}, all flat manifolds are nilmanifolds, since all Abelian groups are nilpotent.
In a sense, nilmanifolds are ``straightforward" generalisations of the flat ones. This can also be seen through the ascending central 
series: nilmanifolds are homeomorphic to torus bundles of torus bundles etc over a torus. From such  topological viewpoint it also becomes 
clear that nilmanifolds are relatively ``simple", but non-trivial, generalisations of flat manifolds as the compact cover of the latter are just tori 
according to Bieberbach's theorem. In the non-simply connected case, it was proved that a closed (compact without boundary) 
aspherical (i.e. having \ $\pi_i, \ i\geq 2$ \ trivial) manifold of dimension greater than $4$,  
whose fundamental group is virtually nilpotent is determined up to homeomorhism by its fundamental group [29], [30]. This is a very 
non-trivial statement when one considers that the asphericity should be just a homotopy invariant rather than a 
homeomorphism-invariant property of a manifold. Indeed, from a topological viewpoint two aspherical manifolds 
(actually CW complexes) are homotopy equivalent if and only if their fundamental groups are isomorphic. Equivalence under 
homeomorphisms  is a far stronger condition and indicates a far greater rigidity that nilmanifolds possess than may originally have 
been suspected. This can be traced back to their asphericity and also be see nas a strengthening of the Borel conjecture [31], [32] 
which claims that if the fundamental groups are of two closed aspherical manifolds (of dimension greater than 4) are homeomorphic 
any homotopy equivalnce is homotopic to a homeomorphism. This issue will be briefly revisited in subsection {\bf 2.8} below.  \\
  

\noindent {\bf 2.6} \ Since according to the Gromov/Ruh theorem mentioned in the previous subsection almost flat manifolds are 
nilmanifolds, results applying to the latter also apply to the former. One implication of interest to the present work, which was 
previously used in [21] in the context of Tsallis entropy is the following: a discrete group is virtually nilpotent if and only if it has polynomial 
growth [33]-[42]. Since in our case we have a nilpotent group \ $G_{\mathbb{R}}$ \ over \ $\mathbb{R}$, \ to apply the theorem we need a 
discrete version of it. To achieve this, we consider the corresponding nilpotent group \ $G_{\mathbb{Z}}$ \ defined over the integers \ 
$\mathbb{Z}$. \ In the limit that the distance between consecutive integers approaches zero, then we recover \ $G_{\mathbb{R}}$. \
This idea can be formalised through the structure of asymptotic cones  for precise definition, properties and examples of which we refer the 
reader to [43], [44]. The results of [22], [23] seem to indicate that the if a system has a configuration or phase space volume that grows
polynomially as a function of a radius, then such systems are constrained enough and sufficiently far from being ergodic that the Tsallis
entropy might be more appropriate than the BGS entropy in describing their statistical behavior. Whether this is indeed true, or what 
additional requirements may have to be met in order for a system to be described by the Tsallis entropy is still unclear [6], at least 
to the best of our knowledge.\\


\noindent{\bf 2.7} \ It should be stressed that the results mentioned in subsections {\bf 2.5} and {\bf 2.6}  
hold under two quite strong assumptions. The first of them is that the generalized addition (7) is realized on the configuration or phase 
space of a system. As was pointed out above, the difference between (7) and the ordinary addition becomes more evident at high values 
of the entropy \ $S_q$. \ In metric terms, this amounts to considering objects at the ideal boundary [45] of the hyperbolic space \ 
$\mathbb{H}^n$. \ This is not completely unexpected: $\mathbb{H}^n$ is a Riemannian manifold and as such it is isometric, to zeroth order,
to the Euclidean space \ $\mathbb{R}^n$ \ [25].  The differences between these two spaces become more pronounced as one 
approaches the ideal boundary of \ $\mathbb{H}^n$. \ Close or at the ideal boundary, one can see, for instance, using the Gauss-Bonnet 
theorem, that all similar triangles are congruent, that \ $\mathbb{H}^n$ \ is not a doubling space, that its geodesics deviate exponentialy 
from each other rather than linearly, that its domains obey a linear rather than power-law isoperimetric inequality etc. 
Moreover, seen from afar, it is evident that the asymptotic cone of \ $\mathbb{R}^n$ \ is \ $\mathbb{R}^n$ \ itself. 
By contrast, the asymptotic cone of \ $\mathbb{H}^n$ \ is an \ $\mathbb{R}$-tree  whose degree has the power of the continuum at every 
point [43], [45] . Hence it is not obvious whether such an effective hyperbolic structure (10) can be realized at the level of the 
configuration/phase space, or whether it is a result of statistical averaging. A case of the latter  are the
limiting distributions arising in probablility theory. Examples are  the Gaussian distribution arising as a result of convergence due to 
the Central Limit theorem or its Gnedenko-Kolmogorov and further extensions to stable (L\'{e}vy) or other distributions. \\          

The second assumption is that  in order to have an almost additive entropy, the above induced hyperbolic structure (10) 
should not only exist  but also be related to an almost vanishing sectional curvature of the configuration/phase space which obeys (25). 
Let's try to illustrate this non-trivial point with a ``non-example". 
The Hamiltonian Mean Field  (HMF) model has attracted considerable attention since its 
introduction [46], [47]. This is due  to the rich and, at times unexpected, behavior that it describes despite its apparent simplicity. 
It can be interpreted as describing \ $N$ \ unit mass particles moving on a unit circle internacting with an infinite range attractive ($J>0$) or 
repulsive ($J<0$) potential. Let \ $\theta_i, \ i=1, \ldots, n$ \ indicate the position of
each of these particles on the unit circle and \ $p_i, \ i=1, \ldots, n$ \ their conjugate momenta. The Hamiltonian of the HMF model in zero 
external magnetic field  
\begin{equation}
    \mathcal{H} \ = \ \frac{1}{2} \sum_{i=1}^N p_i^2 + \frac{J}{2N} \sum_{i,j=1}^N \left[ 1 - \cos (\theta_i - \theta_j) \right]
\end{equation} 
Since this Hamiltonian allows a separate treatment of the angles and their conjugate (angular) momenta in the partition function, we 
can integrate out the latter and focus only on the former in undertanding the dynamical behavior of the system. This essentially simplifies
the evolution of the system from its phase space to its coordinate space which is an $N$-torus \ $\mathbb{T}^n$. \  
The dynamical behavior of the model on \ $\mathbb{T}^N$ \ can be  can be explored by using the Jacobi variational approach [48]. This 
amounts to choosing an appropriate metric for \ $\mathbb{T}^N$ \ determining its geodesics, their stability properties etc. One such metric is 
the Jacobi metric which is conformally related to the flat Euclidean metric and is given by 
\begin{equation}     
    {\bf g}_{ij} \ = \ 2 \left\{ E -  \frac{J}{2N} \sum_{i,j=1}^N \left[1 - \cos (\theta_i - \theta_j) \right] \right\} \delta_{ij}
\end{equation}
where \ $E$ \ is the total energy of the system and \ $\delta_{ij}$ \ is the Kronecker symbol. Using (22), (23) and (24) one can calculate 
the curvature of this Jacobi metric on \ $\mathbb{T}^N$ \ and with some additional assumptions [48] one can obtain the largest Lyapunov 
exponent of the system [49]. Notice that the calculation of [49] uses the Eisenhart metric [48] instead of the Jacobi metric used above, 
and for this reason, it is not  clear to us to what extent the conclusions about the dynamics of the model would agree with the results 
that can arise by using the Jacobi metric. We see from (32)
that for very large values of the total energy \ $E$ \ the metric is almost flat. This corresponds to the particles 
having very high kinetic energies. As the interaction between them is bounded both from above and below, the kinetic energy induces 
an almost Euclidean metric on \ $\mathbb{T}^n$. \ Someone could make the same statement for a total energy much larger then the 
potential energy of a system as long as the latter is bounded from both above and below. 
In this regime, deviation from almost flatness may appear in the thermodynamic limit 
\ $N\rightarrow\infty$. \ One can also lower the energy \ $E$ \ and observe deviations from  (25). Geometrically, one can use \ $E$ \ as a 
control parameter in defining the asymptotic cones of \ $\mathbb{T}^N$ \ endowed with (25). Since \ $\mathbb{T}^N$ \ is compact all 
asymptotic cones will be points, so this line of search does not seem to lead to  any non-trivial results. \\

Assume now that the Hamiltonian HMF model is described by the Tsalis entropy with nonextensve parameter \ $q\in [0,1)$ \ as has been 
assumed throughout this work. We then have to find a hyperbolic analogue of (32).  Assume, as it 
usually tacitly done in most physical applications, that we are interested only in smooth Riemannian metrics on \ $\mathbb{T}^N$. \ 
Putting a metric of negative sectional curvature will decrease [25] the possible number of conjugate points along any 
given geodesic. Therefore, we are really interested in metrics on \ $\mathbb{T}^N$ \ not having any conjugate points. In response to a 
of E. Hopf who settled this for the $N=2$ case [50], it was proved for any $N$  in [51] that all such metrics are flat. Therefore, this formalism
does not seem to be able to decide whether the HMF model should be more effectively be described by \ $S_{BGS}$ \ or by \ $S_q$ \ 
for \ $q\approx 1$ \ for \ $E\rightarrow \infty$, \ before considering the thermodynamic limit \ $N\rightarrow \infty$. \\   
      
         \hspace{5mm}
      

\noindent{\bf 2.8} \ An issue that may be of interest at this point is the following. Suppose that a system is indeed described by the Tsallis 
entropy and that the assumptions of \ {\bf 2.7} \ are valid. We would like to see geometrically how such a system, assumed to be out of 
equilibrium,  could be  coupled to a thermostat, that may be used to bring the system to equilibrium. 
Several proposals exist, especially in the context of molecular dynamics 
simulations. Here we follow the approach, in spirit, if not necessarily in letter, proposed by Nos\'{e} [52] and further developed by 
Hoover [53]. The goal of the Nos\'{e}-Hoover approach was to be able to reproduce  the canonical phase 
space distribution in a way that the kinetic energy can fluctuate following the Maxwell-Boltzmann distribution. Similar arguments
could conceivably be made for an underlying distribution, such as a $q$-Gaussian one [6] associated to the Tsallis entropy. 
In this approach a heat bath is introduced in the Hamiltonian of the system through an additional degree of freedom \ $s$ \ 
its conjugate momentum \ $p_s$ \ and an additional parameter \ $Q$. \ The Hamiltonian proposed [53] for  a system with \ $N$ \ 
degrees of freedom is 
\begin{equation}    
   \mathcal{H}_{NH} \ = \  \sum_{i=1}^N \frac{p_i^2}{2s^2} + U(q_1, \ldots , q_N) + (N+1) k_B T \ln s + \frac{p_s^2}{2Q}
\end{equation}
As can be seen, \ $s$ \ is a time-scale variable and \ $U$ \ indicates the potential energy of the system.
It was proved [52] that the microcanonical distribution of 
\ $(q_i, p_i, s, p_s), \ i=1, \ldots, N$ \ is equivalent to a canonical distribution defined over \ $(q_i, \frac{p_i}{s}), \ i=1, \ldots, N$. \ 
In the context of the present geometric considerations, we  see that the canonical momentum \ $p_s$ \ splits from its conjugate variable
$s$, so it can be integrated out in the canonical partition function. Therefore one has to augment the coordinate space \ $M$ \ 
to incorporate \ $s$ \ in this description. \\

Probably the simplest way to implement this addition of one degre of freedom is by considering the augmented coordinate 
space to be the trivial fiber bundle \ $\bar{M} = M \times\mathbb{R}$. \  Since we would like the formalism to be flexible enough 
to accommodate thermostats that may be defined in a similar manner to that of Nos\'{e}-Hoover by an additional parameter, a 
slightly more general construction may be beneficial. One such case arises if we require \ $M$ \ to be the (topological) boundary of the 
augmented space \ $\bar{M}$, \ indicated by \ $M = \partial\bar{M}$, \  
and additionally demand that the metric \ ${\bf g}$ \ of \ $M$ \ is the induced metric arising from the metric \ ${\bf \bar{g}}$ \ of \ $\bar{M}$ \ 
via the boundary map \ $\partial$. \ A second such construction relies on geometry and arises if we demand \ $M$ \ to be a totally 
geodesic submanifold of the augmented \ $\bar{M}$ \ which should be assumed closed, for simplicity.\\  

In the first case, we can proceed with analogy to the construction leading to (10). What was done in [16] 
was to ``augment" \ $\mathbb{R}$ \ to \ $\mathbb{R}^2$ \ for comparison purposes. This can be extended in the case of \ $M$ \ 
by considering  \ $\bar{M} = M\times (0,1)$ \ which can also topologically seen as a double cone over \ $M$ \ with the apex removed. 
Following (10), the metric on $\bar{M}$ would be 
\begin{equation}
          {\bf \bar{g}} \  = \ du^2 + e^{-2tu} {\bf g}
\end{equation}
The question on whether such a construction is possible, at hte topological level, 
was partially addressed by [54]. It was proved initially by Thom [55] 
that  \ $M = \partial \bar{M}$ \ if all its Stiefel-Whitney classes [56] vanish. More specifically, within the oriented category, Wall [57] 
proved that an orientable manifold is an oriented boundary, if in addition, all its Pontryagin classes [56] vanish. Now, since a nil-manifold 
is parallelizable all its Pontryagin classes vanish too. In [54] it is partially proved that if \ $M$ \ is almost flat then there is a compact smooth 
manifold \ $\bar{M}$ \ such that \ $M = \partial \bar{M}$. \ For the case of our interest the results of [54] 
are of sufficient generality: consider \ $M$ \ to be aspherical with a nilpotent fundamental group \ $\pi_1(M)$. \ 
Then, there is indeed a compact smooth manifold \ $\bar{M}$ \ such that \ $\partial \bar{M} = M$. \
At the metric level, and as long as someone is willing to accept some kind of singularities: it was proved in [58] that every closed flat 
manifold is the boundary of a smooth 
manifold. The corresponding conjecture for almost flat manifolds was put forth in [54], [59].  This conjecture is equivalent to demanding that 
\ $\bar{M}$ \ is a negatively curved manifold. This conjecture proved to be true in the case that \ $M$ \ itself is a nimanifold. Therefore it is 
known that for \ $M$ \ closed, connected and almost flat there is a \ $\bar{M}$ \ which is complete, has finite volume and has negative 
pinched sectional curvature, namely its sectional curvature has the uniform bounds \  $k \in [-k_1, - k_2], \  0 < k_2 \leq k_1 < \infty$.
Moreover, and extending the idea of a metric suspension with both apexes removed, 
such a \ $\bar{M}$ \ has exactly two connected cusps, each of which is diffeomorphic to \ $M \times [k_2, \infty)$. \ In such a case the 
work of [60], [61] essentially proves that the metric \ ${\bf \bar{g}}$ \ of \ $\bar{M}$ \ is a warped cusp metric, namely it has the 
form of (34) with \ $u < u_0 \in \mathbb{R}$. \ This should be sufficiently general for answering in the affirmative our question. 
For further information, an overview and constructions with emphasis on the smooth category, see [62].\\

In the second case, one is interested in a non-trivial \ $\bar{M}$ \ having \ $M$ \ as a totally geodesic submanifold, extending in a 
different way  the trivial bundle structure encountered above. Moreover, to reach non-trivial results one also requires \ $\bar{M}$ \ not  
be splitting as a product. This amounts to having geometric rank 1 and moreover having non-positive sectional curvature \ $k\leq 0$ \
[62], [63]. If such an \ $M$ \ is closed (compact,  without a boundary), then there exist [63] a closed geometric rank 1 manifold \ $\bar{M}$, \ 
having \ $k(\bar{M}) \leq 0$, \ such that there is an isometric embedding \ $M \rightarrow \bar{M}$. \  The negative curvature condition is 
expecially pertinent in the face of (12). Moreover it guarantees that the geodesic flow on \ $(\bar{M}, {\bf \bar{g}})$ \ is still Anosov, so it can 
be used as a prototype metric for systems obeying the chaotic hypothesis [64]. Such systems are thought to be sufficiently 
well-desrcibed by \ $S_{BGS}$ \ however, so the role of the Tsallis entropy among their general class, if any, is unclear. 
Such statements would be much more useful to our purposes, if any conclusions could be drawn about manifolds of variable sign 
sectional curvature, even if bounded both above and below. Regrettably, we are unaware of any such results that could be more pertinent 
to the underlying dynamics whose statistics is described by the Tsallis entropy.\\

                                                                                                    \vspace{8mm}


                                                             \centerline{\large\sc  3. \  \   Conclusions and discussion}              

                                                                                                \vspace{5mm}
    
In th present work we discussed the dynamical underpinnings in a geometric formalism of systems that are almost additive, namely they 
are described by the Tsallis entropy for a nonextensive parameter \ $q\approx 1$. \ We found that a nilpotent structure arises once more,
as in [21], but for entirely different reasons. In contrast to [21] there was no embedding into a more general structure, but the nilpotent 
structure arose as a geometric consequence of the Gromov/Ruh theorem. As a result, we saw once more that the volume of the 
configuration/phase space of systems described by the Tsallis entropy should have a power law growth as was first pointed out in [22], 
[23] and was subsequently argued in [21]. Some topological and metric ideas with potential statistical mechancal relevance 
were discussed in this context. Such ideas and techniques, is hoped, may contribute to a better understanding of the 
dynamical foundations underlying the Tsallis entropy.\\

The approach followed in the present work  is indeed  limited to almost additive systems, where the nonextensive parameter 
\ $q\approx 1$ \ in the Tsallis entropy \ $S_q$. \ Despite this obvious limitation, we see that the emergence of an 
underlying hyperbolicity as expressed in (10), (12) may  indeed be a real conseqence having a dynamical basis, 
rather than just being an artifact of the particular formalism employed. The exact relationship between the nilpotent structures 
encountered here as well as in [21] and hyperbolicity, as suggested by (9), (10), is intimately related to the concepts and techniques 
related to Mostow rigidity [65], [66]. \\

Another point that may be worth exploring in the future is the exact meaning of the word 
``hyperbolicity". Several proposals have been presented in [45], [43] 
about how such a concept could be realized [62], [67]. Which one of them, if any, is applicable in describing the effective dynamics 
underlying the systems whose statistical behavior is encoded by the Tsallis entropy will be the main topic of a future work.  \\

 
 
                                                                              \centerline{\large\sc Acknowledgement}    
 
                                                                                                 \vspace{3mm}
 
                            \noindent    We are grateful to Professor C. Tsallis for his comments on the manuscript.\\
 
                                                                                              \vspace{8mm}
 
                                                                                                                         
                                                        \centerline{\large\sc References}
 
                                                                          \vspace{5mm}

\noindent [1] J. Harvda, F. Charvat, \ \emph{Kybernetica} {\bf 3}, \ 30 \ (1967).\\
\noindent [2] Z. Dar\'{o}czy,  \ \emph{Inf. Comp. / Inf. Contr.} {\bf 16}, \ 36 \ (1970).\\
\noindent [3] N.A. Cressie, T.R. Read, \ \emph{J. R. Stat. Soc. B}{\bf 46}, \ 440 \ (1984).\\
\noindent [4] T.R. Read, N.A. Cressie, \ \emph{Goodness of Fit Statistics for Discrete Multivariate Data}, \ Springer \\
                            \hspace*{4mm} (1988).\\
\noindent [5] C. Tsallis, \ \emph{J. Stat. Phys.} {\bf 52}, \ 479 \ (1988).\\
\noindent [6] C. Tsallis, \ \emph{Introduction to Nonextensive Statistical Mechanics: Approaching a Complex \\ 
                           \hspace*{4mm} World}, \  Springer \  (2009).\\
\noindent [7] C.E. Shannon, \ \emph{Bell Syst. Tech. J.} {\bf 27}, \ 379 \ (1948).\\
\noindent [8] C.E. Shannon, \ \emph{Bell Syst. Tech. J.} {\bf 27}, \ 623 \ (1948).\\
\noindent [9] A.J. Khintchin, \  \emph{Ups. Mat. Nauk.} {\bf 8}, \ 3 \ (1953).\\ 
\noindent [10] R.J.V. Santos, \ \emph{J. Math. Phys.} {\bf 38}, \ 4104 \ (1997).\\
\noindent [11] S. Abe, \ \emph{Phys. Lett. A} {\bf 271}, \ 74 \ (2000).\\
\noindent [12] L. Nivanen, A. Le Mehaut\'{e}, Q.A. Wang, \ \emph{Rep. Math. Phys.} {\bf 52}, \ 437 \ (2003).\\
\noindent [13] E.P. Borges, \ \emph{Physica A} {\bf 340}, \ 95 \ (2004).\\
\noindent [14] T.C. Petit Lob\~{a}o, P.G.S. Cardoso, S.T.R. Pinho, E.P. Borges, \ \emph{Braz. J. Phys.} {\bf 39}, \ 402 \\
                           \hspace*{7mm} (2009).\\
\noindent [15] N. Kalogeropoulos, \ \emph{Physica A} {\bf 391}, \ 1120 \ (2012).\\
\noindent [16] N. Kalogeropoulos, \ \emph{Physica A} {\bf 391}, \ 3435 \ (2012).\\
\noindent [17] N. Kalogeropoulos, \ \emph{QScience Connect} {v2012.\bf 12}\\
\noindent [18] N. Kalogeropoulos, \ \emph{Vanishing largest Lyapunov exponent and Tsallis entropy},\\
                             \hspace*{7mm} {\sf arXiv:1203.2707}\\ 
\noindent [19] N. Kalogeropoulos, \ \emph{Escort distributions and Tsallis entropy}, {\sf arXiv: 1206.2707}\\
\noindent [20] A.J. Creaco, N. Kalogeropoulos, \ \emph{J. Phys. Conf. Ser.} {\bf 410}, \ 012148 \ (2013).\\
\noindent [21] N. Kalogeropoulos, \ \emph{Tsallis entropy composition and the Heisenberg group}, \ to be published \\
                             \hspace*{7mm} in \  \emph{Int. J. Geom. Meth. Mod. Phys.} \ (August 2013), \ {\sf arXiv:1301.0069}\\ 
\noindent [22] C. Tsallis, M. Gell-Mann, Y. Sato, \ \emph{Proc. Nat. Acad. Sci.} {\bf 102}, \ 15377 \ (2005).\\
\noindent [23] R. Hanel, S. Thurner, \ \emph{Europhys. Lett.} {\bf 96}, \ 50003 \ (2011).\\
\noindent [24] P. Petersen, \ \emph{Riemannian Geometry}, \ 2nd Edition, \ Springer \ (2006).\\ 
\noindent [25] M. Gromov, \ \emph{Rondi. Sem. Mat. Phys. Milano} \ {\bf 61}, \ 9 \ (1991).\\
\noindent [26] M. Gromov, \ \emph{J. Diff. Geom.} {\bf 13}, \ 231 \ (1978).\\
\noindent [27] P. Buser, H. Karcher, \ \emph{Asterique} {81}, \ 1 \ (1981).\\
\noindent [28] E. Ruh, \ \emph{J. Diff. Geom.} {\bf 17}, \ 1 \  (1982).\\
\noindent [29] F.T. Farrell, W.C. Hsiang, \ \emph{Amer. Jour. Math.} {\bf 105}, \ 641 \ (1983).\\ 
\noindent [30] F.T. Farrell, L.E. Jones, \ \emph{Topological rigidity for compact non-positively curved manifolds}, \\
                              \hspace*{7mm} in \ \emph{Differential Geometry: Riemannian Geometry (Los Angeles (1990))}, \  Amer. Math.\\
                              \hspace*{7mm}   Soc. \  (1993).\\
\noindent [31] F.T. Farrell, \emph{The Borel Conjecture}, in \emph{High-Dimensional Manifold Topology, Trieste 2001}, \\
                             \hspace*{7mm} available at \ \ {\sf users.ictp.it/ $\widetilde{}$ pub\_off/lectures/lns009/Farrell/Farrell.pdf}\\
\noindent [32] M. Kreck, W. L\"{u}ck, \emph{The Novikov Conjecture: Geometry and Algebra}, \ Birkh\"{a}user (2005).\\ 
\noindent [33] A. \v{S}var\v{c}, \ \emph{Dokl. Acad. Nauk. SSSR} {\bf 105}, \ 32 \ (1955).\\
\noindent [34] V. Efremovich, \ \emph{Usp. Mat. Nauk.} {\bf 8}, \ 189 \ (1953).\\
\noindent [35] J.A.  Wolf, \ \emph{J. Diff. Geom.} {\bf 2}, \ 421 \ (1968).\\
\noindent [36] J. Milnor, \ \emph{J. Diff. Geom.} {\bf 2}, \ 447 \ (1968).\\
\noindent [37] Y. Guivarc'h, \ \emph{C.R. Acad. Sci. Par.} {\bf 272}, \ 1695 \ (1971).\\ 
\noindent [38] H. Bass, \ \emph{Proc. Lond. Math. Soc. (3)} {\bf 25}, \ 603 \ (1972).\\ 
\noindent [39] M. Gromov, \ \emph{Publ. Math. I.H.E.S.} {\bf 53}, \ 53 \ (1981).\\
\noindent [40] P. Pansu, \ \emph{Ergod. Th. \& Dynam. Syst.} {\bf 3}, \ 415 \ (1983).\\
\noindent [41] B. Kleiner, \ \emph{J. Amer. Math. Soc.} {\bf 23}, \ 815 \ (2010).\\
\noindent [42] Y. Shalom, T. Tao, \ \emph{Geom. Funct. Anal.} {\bf 20}, \ 1502 \ (2010). \\
\noindent [43] M. Gromov, \emph{Asymptotic Invariants of Infinite Groups}   in  \emph{Geometric Group Theory, Vol.2}\\
                             \hspace*{7mm} G.A. Niblo, M.A. Roller (Eds.), \ Cambridge University Press \ (2003).\\
\noindent [44] C. Drutu, \ \emph{Int. J. Algebra Comput.} {\bf 12}, \ 99 \ (2002).\\
\noindent [45] M. Gromov, \emph{Hyperbolic groups},  in  \emph{Essays in group theory}, S. Gersten (Ed.), MSRI Publ. {\bf 8}\\
                              \hspace*{7mm} Springer \ (1987).\\
\noindent [46] A. Campa, T. Dauxois, S. Ruffo, \ \emph{Phys. Rep.} {\bf 480}, \ 57 \ (2009).\\ 
\noindent [47] T. Dauxois, V. Latora, A. Rapisarda, S. Ruffo, A. Torcini, \emph{The Hamitonian Mean Field \\
                             \hspace*{7mm} Model: from Dynamics to Statistical Mechanics and back}, \ in \ \emph{Dynamics and \\
                             \hspace*{7mm}  Thermodynamics of Systems with Long Range Interactions}, \ T. Dauxois, S. Ruffo,\\
                             \hspace*{7mm} E. Arimondo, M. Wilkens \ (Eds.), \ Lect. Notes. Phys. Vol. {\bf 602}, \ Springer \ (2002).\\  
\noindent [48] L. Casetti, M. Pettini, E.G.D. Cohen, \ \emph{Phys. Rep.} {\bf 337}, \ 237 \ (2000).\\
\noindent [49] M.-C. Firpo, \ \emph{Phys. Rev. E} {\bf 57}, \  6599 \ (1998).\\
\noindent [50] E. Hopf, \ \emph{Proc. Nat. Acad. Sci.} {\bf 34}, (1948).\\
\noindent [51] D. Burago, S. Ivanov, \ \emph{Geom. Funct. Anal.} {\bf 4}, \ 259 \ (1994).\\
\noindent [52] S. Nos\'{e}, \ \emph{J. Chem. Phys.} {\bf 81}, \ 511 \ (1984).\\  
\noindent [53] W.G. Hoover, \ \emph{Phys. Rev. A} {\bf 31}, \ 1695 \ (1985).\\
\noindent [54] F.T. Farrell, S. Zdravkovska, \ \emph{Michigan Math. J.} {\bf 30}, \ 199, (1983).\\
\noindent [55] R. Thom, \ \emph{Comment. Math. Helv.} {\bf 28}, \ 17 \ (1954).\\ 
\noindent [56] J. Milnor, J. Stasheff, \ \emph{Characteristic Classes}, \ Princeton University Press \ (1974).\\
\noindent [57] C.T.C. Wall, \ \emph{Surgery on Compact Manifolds}, \ LMS Vol. 1, \ Academic Press \ (1970).\\ 
\noindent [58] G.C. Hamrick, D.C. Royster, \ \emph{Invent. Math.} {\bf 66}, \ 405 \ (1982).\\
\noindent [59] S.-T. Yau, \ \emph{Open problems in differential geometry}, \  in  \emph{Proc. Symp. Pure Math.} {\bf 54} \ (1993).\\
\noindent [60] Z.M. Shen, \ \emph{Pacific  J. Math.} {\bf 163},  \ 175 \ (1994).\\
\noindent [61] I. Belegradek, V. Kapovich, \ \emph{Acta Math.} {\bf 196}, \ 229 \ (2006).\\
\noindent [62] P. Ontaneda, \ \emph{Pinched Smooth Hyperbolization}, \ {\sf arXiv:1110.6374}\\
\noindent [63] T. T\^{a}m Nguy{\^{e}}n Phan, \ \emph{Nonpositively Curved Manfolds Containing a Prescribed\\
                             \hspace*{7mm}  Nonpositively Curved Hypersurface}, \ {\sf arXiv:1206.0098}\\  
\noindent [64] G. Gallavotti, E.G.D. Cohen, \ \emph{J. Stat. Phys.} {\bf 80}, \ 931 \ (1995).\\
\noindent [65] G.D. Mostow, \ \emph{Strong Rigidity of Locally Symmetric Spaces}, \ Princeton University Press  \\
                              \hspace*{7mm}  (1973).\\
\noindent [66] P. Pansu, \ \emph{Ann. Math.} {\bf 129}, \ 1 \  (1989).\\
\noindent [67] R.M. Charney, M.W. Davies, \ \emph{Topology} {\bf 34}, \ 329 \ (1995).\\

\end{document}